\documentstyle[fleqn,espcrc2,epsf]{article}
\title{
$I=2$ Pion Scattering Length with Wilson Fermions
\thanks{presented by N. Ishizuka}
}
\author{
JLQCD Collaboration : 
S.~Aoki
\address{
Institute of Physics, University of Tsukuba,
Tsukuba, Ibaraki 305-8571, Japan
},
M.~Fukugita
\address{
Institute for Cosmic Ray Research,
University of Tokyo,
Tanashi, Tokyo 188-8502, Japan
},
S.~Hashimoto
\address{
High Energy Accelerator Research Organization (KEK),
Tsukuba, Ibaraki 305-0801, Japan
},
K-I.~Ishikawa$^{\rm ~c}$,
N.~Ishizuka$^{\rm ~a,}$\address{
Center for Computational Physics,
University of Tsukuba,
Tsukuba, Ibaraki 305-8577, Japan
},
Y.~Iwasaki$^{\rm ~a,d}$
K.~Kanaya$^{\rm ~a,d}$,
T.~Kaneda$^{\rm ~a}$,
S.~Kaya$^{\rm ~c}$,
Y.~Kuramashi$^{\rm ~c}$,
M.~Okawa$^{\rm ~c}$,
T.~Onogi
\address{
Department of Physics, Hiroshima University,
Higashi-Hiroshima, Hiroshima 739-8526, Japan
},
S.~Tominaga$^{\rm ~c}$,
N.~Tsutsui$^{\rm ~e}$,
A.~Ukawa$^{\rm ~a,d}$,
N.~Yamada$^{\rm ~e}$,
T.~Yoshi\'{e}$^{\rm ~a,d}$
}
\begin{document}
%
%
\begin{abstract}
We present results for $I=2$ pion scattering length
with the Wilson fermions in the quenched approximation.
The finite size method presented by L\"uscher is employed,
and calculations are carried out
at $\beta=5.9$, $6.1$, and $6.3$.
In the continuum limit, we obtain a result in reasonable
agreement with the experimental value.
\end{abstract}
\maketitle
%
%
\section{ Introduction }
Lattice calculations of scattering lengths of the two-pion system
is an important step for understanding of strong interactions
beyond the hadron mass spectrum.
For the $I=0$ process, which is difficult
due to the presence of disconnected contributions,
only one group carried out the calculation~\cite{Kuramashi}.
For the $I=2$ process, on the other hand,
a number of calculations has been carried out with the Staggered~\cite{Kuramashi,SGK}
and the Wilson fermions~\cite{Kuramashi,GPS}.
These calculations reported results in agreement with the prediction of current
algebra or lowest-order chiral perturbation theory(CHPT)~\cite{Winberg}.
It is known, however,
that this prediction differs from the experimental value over $1.4\sigma$,
and that the higher order effects of CHPT
are small~\cite{Gasser-Leutwyler:Bijinens}.
%
%
%
%
%
%
%

The past lattice calculations were made on coarse lattices with small sizes,
and the continuum extrapolation was not taken.
In this article we report on our high statistics calculation
of the $I=2$ pion scattering length aiming to improve on these points.
This work is carried out in quenched lattice QCD employing
the standard plaquette action for gluons
with the Wilson fermions.
The number of configurations (and lattice size) are
$187 (16^3 \times 64 )$,
$120 (24^3 \times 64 )$, and
$100 (32^3 \times 80 )$ for $\beta=5.9$, $6.1$, and $6.3$.
The pion mass covers the range $450 - 900{\rm MeV}$.
%
%
\section{ Method }
The energy eigenvalue of a two-pion system in a finite periodic box $L^3$
is shifted by finite-size effect.
L\"uscher presented a relation between the energy shift $\Delta E$
and the $S$-wave scattering length $a_0$ given by~\cite{Lusher}
\begin{equation}
   - \Delta E \cdot  \frac{ m_\pi L^2 }{ 4 \pi^2 }
= T + A \cdot T^2 + B \cdot T^3 + O( T^4 ) \ ,
\label{Lusher.eq}
\end{equation}
where $T = a_0 / ( \pi L )$.
Since $T$ takes a small value, typically $\sim - 10^{-2}$, in our simulation
we can neglect the higher order terms $O(T^4)$.
The constants $A$ and $B$ are geometrical values
$A = -8.9136 \cdots $ and $ B = 62.9205\cdots $.

The energy shift $\Delta E$ can be obtained from the ratio
$R( t ) = G( t )/D( t )$, where
\begin{eqnarray}
&&   G( t ) = \langle \pi^+(t) \pi^+ (t) W^{-}(t_1) W^{-}(t_2 ) \rangle  \cr
&&   D( t ) = \langle \pi^+(t) W^{-} (t_1)  \rangle \
              \langle \pi^+(t) W^{-} (t_2)  \rangle \ .
\end{eqnarray}
In order to enhance signals we use wall sources (denoted by $W^{-}$)
and fix gauge configurations to the Coulomb gauge.
The two wall sources are placed at different time slices $t_1$ and $t_2$
to avoid contaminations from Fierz-rearranged terms in the two-pion state
that would occur for the choice $t_1=t_2$.
In this work we set $t_2=t_1+1$ and
$t_1=8$, $10$, $13$ for $\beta=5.9$, $6.1$, $6.3$.
Quark propagators are solved with the Dirichlet boundary condition
imposed in the time direction
and the periodic boundary condition in the space directions.
In region $t >> t_1, t_2$ the ratio behaves as
$R(t) \sim Z ( 1  - \Delta E \cdot ( t - t_1 ) + O(t^2) )$.

\begin{figure}[t]
\vspace*{-0.6cm}
\centerline{\epsfxsize=7.5cm \epsfbox{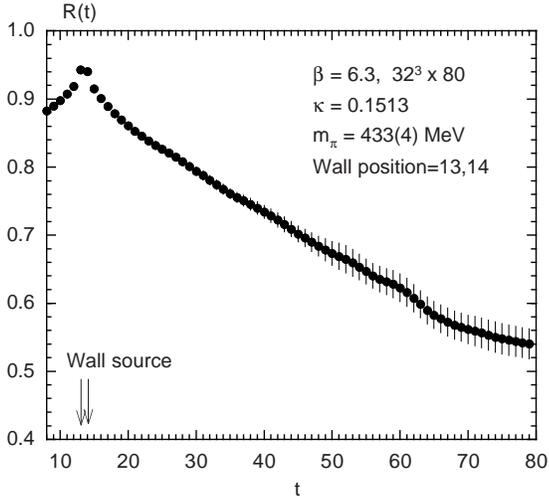}}
\vspace*{-1.0cm}
\caption{\label{L.00.TR.fig}
$R(t)=G(t)/D(t)$.
}
\vspace*{-0.6cm}
\end{figure}

As an example, the ratio $R(t)$ at $\beta=6.3$ and $\kappa =0.1513$
corresponding to $m_\pi= 433(4){\rm MeV}$ is plotted in Fig.~\ref{L.00.TR.fig}.
The signal is very clear and $Z\sim 1$.
This means the overlap of our wall sources with the two-pion state
is sufficiently large.
In general there are higher order terms $O(t^2)$ in $R(t)$,
but we cannot resolve them in Fig.~\ref{L.00.TR.fig}.
Making a linear fitting in the range $t=27-62$,
we obtain $( a \Delta E ) = 5.97(60)\times 10^{-3}$.
Solving equation (\ref{Lusher.eq}) with this value
we obtain $T=-1.73(15)\times 10^{-2}$,
which corresponds to $a_0 = -0.525(45) (1/{\rm GeV})$.
%
%
%
%
%
\section{ Result }
\begin{figure}[t]
\vspace*{-0.6cm}
\centerline{\epsfxsize=7.5cm \epsfbox{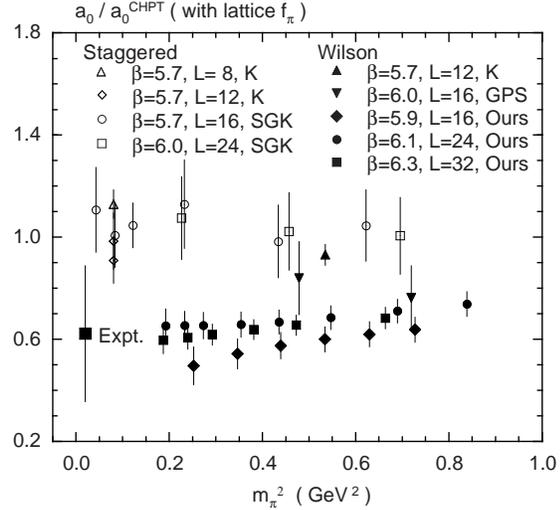}}
\vspace*{-1.0cm}
\caption{\label{comp.RCHPT.fig}
Comparison of our results with the old calculations by $a_0/a_0^{\rm CHPT}$.
}
\vspace*{-0.6cm}
\end{figure}

In Fig.~\ref{comp.RCHPT.fig}
we compare our new results for $a_0 / a_0^{\rm CHPT}$
with those of old calculations,
where $a_0^{\rm CHPT}$ is the prediction of current algebra :
$a_0^{\rm CHPT} = - m_\pi / ( 16 \pi f_\pi^2 )$.
For the decay constant $f_\pi$ we use the value
at finite $m_\pi$ at finite lattice spacing referred in each study.
This ratio has been commonly employed to make a comparison of current algebra
and lattice calculations with different quark actions and parameters.
The open symbols refer to results with the Staggered fermions
and filled ones are those of the Wilson fermions.
The legends give $\beta$, the spatial lattice size $L$,
and collaborations of the studies
(
K : Kuramashi {\it et.al.}~\cite{Kuramashi},
SGK : Sharpe  {\it et.al.}~\cite{SGK},
GPS : Gupta   {\it et.al.}~\cite{GPS},
Ours : our calculation
).
We also plot the experimental value at $m_\pi = 140{\rm MeV}$.
We find that our results are inconsistent with old results,
especially with those of the Staggered fermions.
%
%
%
%

\begin{figure}[t]
\vspace*{-0.6cm}
\centerline{\epsfxsize=7.5cm \epsfbox{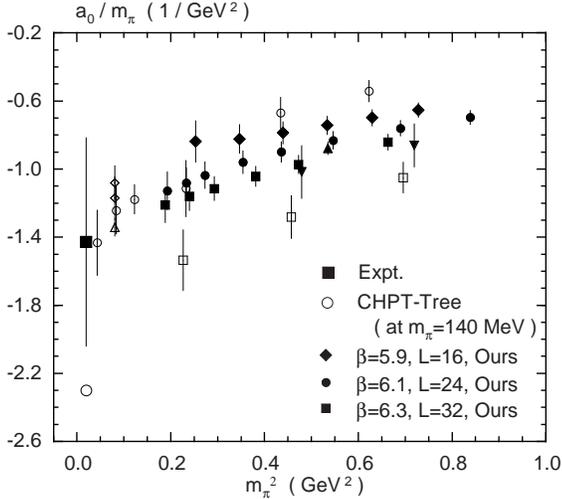}}
\vspace*{-1.0cm}
\caption{\label{comp.a0_mpi.fig}
Comparison of our results with the old calculations by $a_0/m_\pi$.
}
\vspace*{-0.6cm}
\end{figure}

A possible cause of the discrepancy is the systematic error of determination
of $f_\pi$ needed to calculate $a_0^{\rm CHPT}$.
In Fig.~\ref{comp.a0_mpi.fig}
we compare our results with old calculations in terms of $a_0/m_\pi$.
The same symbols as those in Fig.~\ref{comp.RCHPT.fig} are used.
The lattice results including ours are almost consistent with each other.
Also they appear to be in more agreement with the experiment than with
the prediction of current algebra.

We note that the calculation of $f_\pi$, being determined
by the amplitude of correlation function of pion and axial vector current,
is quite difficult.
Various systematic errors may well enter in their determinations.
Further the mass dependence of $f_\pi$ is not small and
$a_0/a_0^{\rm CHPT}$ is very sensitive to it.
For these reasons we analyze $a_0/m_\pi$ below.
%
%
%
%

From chiral symmetry $a_0/m_\pi$ behaves as
\begin{equation}
  a_0/m_\pi = A + B \cdot m_\pi^2 + C \cdot m_\pi^2 \log( m_\pi^2 / \Lambda^2 )
   + O(m_\pi^4) \ .
\end{equation}
For the Wilson fermions we should consider another term
$\propto 1/m_\pi^2$ that arises from breaking of chiral symmetry.
Golterman and Bernard also proposed the same term pointing out that it
can appear from quenching effect~\cite{Golterman-Bernard}.
However, these effects are very small in our simulation as we do not
observe a rapid variation of $a_0/m_\pi$ expected from such a term
in Fig.~\ref{comp.a0_mpi.fig}.
Further the chiral logarithm term, $m_\pi^2 \log( m_\pi^2 / \Lambda^2 )$,
is also small.

\begin{figure}[t]
\vspace*{-0.6cm}
\centerline{\epsfxsize=7.5cm \epsfbox{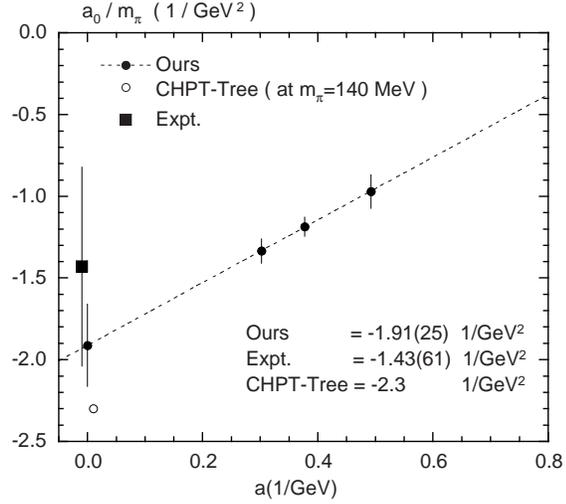}}
\vspace*{-1.0cm}
\caption{\label{a0_mpi.cont_limit.fig}
$a_0/m_\pi$ at the chiral limit at each $\beta$.
}
\vspace*{-0.6cm}
\end{figure}

In Fig.~\ref{a0_mpi.cont_limit.fig} our results for $a_0 / m_\pi$
in the chiral limit obtained by a linear fitting in $m_\pi^2$
are plotted, together with the experimental value
and the prediction of current algebra at $m_\pi=140{\rm MeV}$.
We observe a very clear linear dependence in the lattice spacing $a$.
By a linear extrapolation, we then obtain
\begin{eqnarray}
&& a_0 / m_\pi =  -  1.91(25) \ (1/{\rm GeV^2}) \cr
&& a_0 m_\pi   = - 0.0374(49) \ ,
\end{eqnarray}
in the continuum limit.

This result is consistent with the experimental  value:
$a_0/m_\pi   = -1.43(61)(1/{\rm GeV^2})$ ($a_0 m_\pi = -0.028(12)$).
The difference from the prediction of the current algebra given by
$a_0 / m_\pi = -2.3     (1/{\rm GeV^2})$ ($a_0 m_\pi = -0.045$)
is about $1.5\sigma$.
Since scaling violation is not small, and our data points are far from
the continuum limit, as seen in Fig.~\ref{a0_mpi.cont_limit.fig},
further calculations nearer to the continuum limit is desirable.
In addition studies with the Staggered fermions
should be repeated in a systematic manner
for comparison with present results.
%
%
%
%
\hfill\break

This work is supported by the Supercomputer Project No.45 (FY1999)
of High Energy Accelerator Research Organization (KEK),
and also in part by the Grants-in-Aid of the Ministry of
Education (
Nos.
09304029, 
10640246, 
10640248, 
10740107, 
10740125,
11640294, 
11740162
).
K-I.I is supported by the JSPS Research Fellowship.
%
%
%

%
%
\end{document}